\documentclass[runningheads]{svmult}

\usepackage{makeidx}   
\usepackage{graphicx}  
\usepackage{subeqnar}  
\usepackage{multicol}  
\usepackage{physprbb}  
\makeindex             

\def\lesssim{\mathrel{\hbox{\rlap{\hbox{\lower4pt\hbox{$\sim$}}}\hbox{$<$}}}}
\def\gtrsim{\mathrel{\hbox{\rlap{\hbox{\lower4pt\hbox{$\sim$}}}\hbox{$>$}}}}

\begin{document}
\title*{Light Curves from an Expanding Relativistic Jet}
\titlerunning{Light Curves from an Expanding Relativistic Jet}
%
\author{J. Granot\inst{1}
\and M. Miller\inst{2}
\and T. Piran\inst{1}
\and W.M. Suen\inst{2}
\and P.A. Hughes\inst{3}}
\authorrunning{Jonathan Granot et al.}
%
%
\institute{Racah Institute of Physics, Hebrew University, Jerusalem 91904, Israel
\and Department of Physics, Washington University,
  St. Luis, MO 63130, USA
\and Department of Astronomy, University of Michigan, Ann
  Arbor, MI 48109, USA}

\maketitle              

\begin{abstract}
  \index{abstract} We perform fully relativistic hydrodynamic
  simulations of the deceleration and lateral expansion of a
  relativistic jet as it expands into an ambient medium. The
  hydrodynamic calculations use a 2D adaptive mesh refinement (AMR)
  code, which provides adequate resolution of the thin shell of matter
  behind the shock. We find that the sideways propagation is different
  than predicted by simple analytic models. The physical conditions at
  the sides of the jet are found to be significantly different than at
  the front of the jet, and most of the emission occurs within the
  initial opening angle of the jet.  The light curves, as seen by
  observers at different viewing angles with respect to the jet axis,
  are then calculated assuming synchrotron emission. For an observer
  along the jet axis, we find a sharp achromatic `jet break' in the
  light curve at frequencies above the typical synchrotron frequency,
  at $t_{jet}\approx 5.8(E_{52}/n_1)^{1/3}(\theta_0/0.2)^{8/3}$ days,
  while the temporal decay index $\alpha$ ($F_{\nu}\propto
  t^{\alpha}$) after the break is steeper than $-p$ ($\alpha=-2.85$
  for $p=2.5$). At larger viewing angles $t_{jet}$ increases and the
  jet break becomes smoother.
\end{abstract}

\section{The Hydrodynamics}
\label{HD}
The hydrodynamic simulations are performed using a fully relativistic
2D AMR code. The jet propagates into a cold and homogeneous ambient
medium of number density $n_1\ {\rm cm}^{-3}$. We use $n_1=1$, typical
of the ISM. We assume an adiabatic evolution (i.e. that radiation
losses do not influence the dynamics). For the initial conditions, we use
a wedge of opening angle $\theta_0=0.2$ taken out of the Blandford
McKee \cite{BM} self similar spherical solution, with an isotropic
equivalent energy of $E_{52}10^{52}$ ergs, where we use $E_{52}=1$.
Since the lateral expansion is expected to become significant only
when the Lorentz factor of the flow drops to $\gamma\sim
1/\theta_0=5$, the initial Lorentz factor of the shock was chosen to
be $\Gamma=\sqrt{2}\gamma=23.7$.

Figure (\ref{fig1}) shows a 3D view of the jet at the last time step
of the simulation, with color-maps of the proper number density, $n'$,
and proper emissivity.  Primed quantities are measured in the local
rest frame of the fluid.  While the number density does not change
significantly between the front and the sides of the jet, the
emissivity is significant only at the front of the jet, and drops
sharply at angles larger than the initial opening angle, $\theta_0$.
The overall egg-shaped structure is very different from the
quasi-spherical structure assumed in 1D analytic models.

\section{The Light Curves}
\label{LC}
The synchrotron spectral emissivity, $P'_{\nu'}$, is calculated using
the physical conditions determined by the hydrodynamic simulation. The
electrons and the magnetic field are assumed to hold fractions
$\epsilon_e$ and $\epsilon_B$, respectively, of the internal energy density,
$e'$, while the electrons posses a power law energy distribution,
$N(\gamma_e)\propto \gamma_e^{-p}$ for
$\gamma_e>\gamma_m=\left[(p-2)/(p-1)\right](\epsilon_e e'/n' m_e
c^2)$.  For simplicity, we ignore the effects of electron cooling and
self absorption, so that $P'_{\nu'}\propto\nu'^{1/3}$ for
$\nu'<\nu'_m$ and $P'_{\nu'}\propto\nu'^{(1-p)/2}$ for $\nu'>\nu'_m$,
where $\nu'_m$ is the local synchrotron frequency of an electron with
$\gamma_e=\gamma_m$;  $F_{\nu}$ is calculated using the formalism of
\cite{GPS}, summing over the contributions from the finite 4-volume of the
simulation.

Figure (\ref{fig2}) shows the radio light curves seen by an observer
along the jet axis ($\theta_{obs}=0$). For simplicity, cosmological
corrections are not included. The insert shows an optical light curve
as seen by observers at three different viewing angles with respect to
the jet axis: $\theta_{obs}/\theta_0=0,1,2$. We obtain an achromatic
`jet break' in the light curve for $\nu>\nu_m(t_{jet})$ (as predicted
by simple semi-analytic models \cite{Rhoads,SPH}) at
$t_{jet}(\theta_{obs})$, where $0.66t_{jet}(\theta_0)\approx
t_{jet}(0)\approx 5.8(E_{52}/n_1)^{1/3}(\theta_0/0.2)^{8/3}$ days.
Defining $\alpha$, $\beta$ by $F_{\nu}\propto \nu^{\beta}t^{\alpha}$,
the shape of the break may be approximated by
\begin{equation}\label{shape_of_break}
F_{\nu}=F_0\nu^\beta\left[(t/t_{jet})^{-s\alpha_1}+
(t/t_{jet})^{-s\alpha_2}\right]^{-1/s}\ ,
\end{equation}
where $\alpha_1=\alpha(t\ll t_{jet})$, $\alpha_2=\alpha(t\gg t_{jet})$
and the parameter $s(\theta_{obs})$ determines the sharpness of the
break, and ranges between $s(\theta_0)\approx 1$ to $s(0)\approx 4.5$, 
indicating that the break is sharper at smaller $\theta_{obs}$.
For $\nu<\nu_m(t_{jet})$, there is only a moderate and more
gradual change in $\alpha$, until the time when
$\nu_m=\nu$. 
For $\nu>\nu_m$ we find that $\alpha_2$ is slightly
smaller than $-p$ (the value predicted by most simple models) for
$\theta_{obs}=0$ ($\alpha_2=-2.85$ for $p=2.5$).

\section{Discussion}
\begin{figure}
\begin{center}
\includegraphics[width=6.8cm]{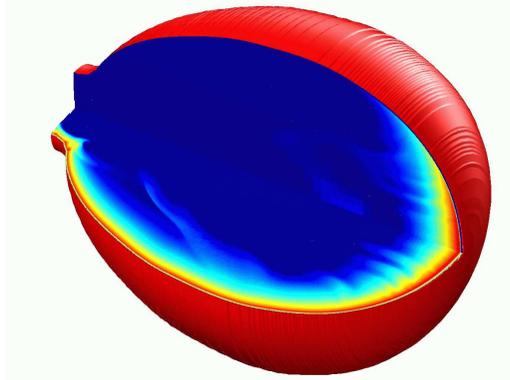}
\end{center}
\caption[]{A 3D view of the jet at the last time step of the simulation. The outer
  surface represents the shock front while the two inner faces show
  the proper number density ({\it lower face}) and proper emissivity
  ({\it upper face}) in a logarithmic color scale}
\label{fig1}
\end{figure}

\begin{figure}
\begin{center}
\includegraphics[width=11cm]{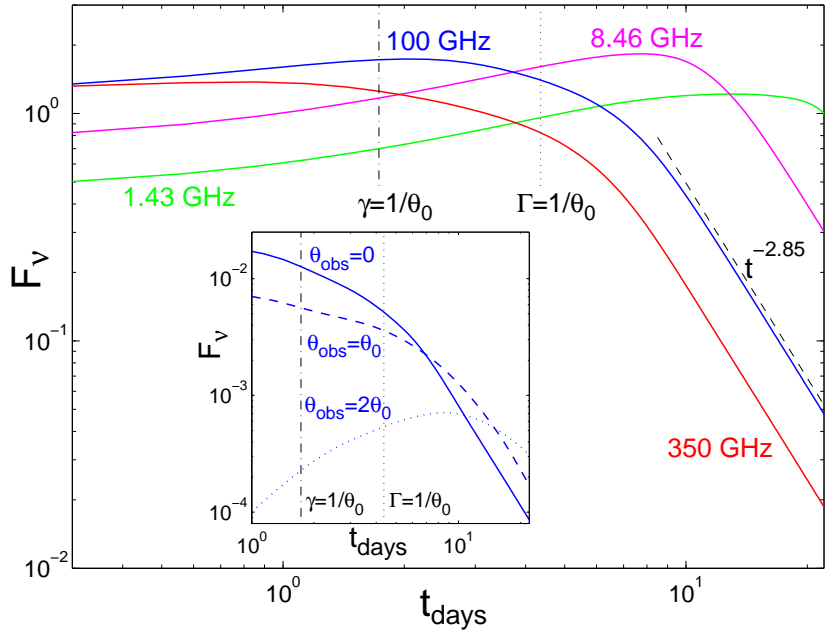}
\end{center}
\caption[]{Radio light curves (flux density, $F_{\nu}$, in arbitrary
  units, as a function of the observed time in days) for an
  observer along the jet axis. We use $\epsilon_e=\epsilon_B=0.1$,
  $p=2.5$. The observed times, for an observer along the jet axis,
  when the Lorentz factors $\gamma$ of the shocked fluid ({\it
    vertical dash-dotted line}) or $\Gamma$ of the shock ({\it
    vertical dotted line}) drop to $1/\theta_0$, for an extrapolated
  spherical evolution, are indicated. ({\bf insert}) Optical light curves
  for observers at viewing angles $\theta_{obs}/\theta_0=0,1,2$
  with respect to the jet axis.}
\label{fig2}
\end{figure}

We find that the physical conditions at the sides of the jet are
significantly different than at the front of the jet, and most of the
radiation is emitted within the initial opening angle of the jet
[$\theta<\theta_0$; see Figure (\ref{fig1})].  Therefore, the
frequently used assumption of a homogeneous jet seems inadequate.  For
$\nu>\nu_m(t_{jet})$ we find a sharp achromatic break in the light
curve at $t_{jet}=5.8(E_{52}/n_1)^{1/3}(\theta_0/0.2)^{8/3}$ days, for
$\theta_{obs}=0$, while at larger
viewing angles $t_{jet}$ increases and the break becomes smoother.
For $\nu<\nu_m(t_{jet})$ the change in the temporal index $\alpha$
near $t_{jet}$ is more moderate and gradual. The value of $\alpha$
for $t>t_{jet}$, $\nu>\nu_m$, $\theta_{obs}=0$ is slightly smaller
than $-p$ ($\alpha=-2.85$ for $p=2.5$). Finally, we note that
$t_{jet}(\theta_{obs}=\theta_0)\approx 1.5 t_{jet}(\theta_{obs}=0)$.
Since, in order to detect a burst in $\gamma$-rays we require that
$\theta_{obs}\lesssim\theta_0$, this may induce an uncertainty of up
to $15\%$ when deducing the value 
of $\theta_0$ from $t_{jet}$, unless
$\theta_{obs}$ is well constrained.

%

\end{document}